\documentclass[letter]{jpsj2} 
\usepackage{color}

\title{Flat-Bands on Partial Line Graphs \\ ---
Systematic Method for Generating Flat-Band Lattice Structures  ---}

\author{
  Shin \textsc{Miyahara}\thanks{E-mail address: miyahara@phys.aoyama.ac.jp}, 
  Kenn \textsc{Kubo},
  Hiroshi \textsc{Ono},
  Yoshihiro \textsc{Shimomura} and
  Nobuo \textsc{Furukawa}
}

\inst{
  Department of Physics and Mathematics,
  Aoyama Gakuin University, Sagamihara 229-8558, Japan
}

\abst{
  We introduce a systematic method for constructing a class of 
  lattice structures that we call ``partial line graphs''.
  In tight-binding models on partial line graphs, 
  energy bands with flat energy dispersions emerge.
  This method can be applied to two- and three-dimensional
  systems. We show examples of partial line graphs
  of square and cubic lattices.
  The method is useful in providing a guideline for
  synthesizing materials with flat energy bands,
  since the tight-binding models on the partial 
  line graphs provide us	
  a large room for modification, maintaining the flat energy dispersions.
}       

\kword{flat-band, partial line graph, tight-binding model, ferromagnetism}

\begin{document}
\maketitle


It has been known that several novel phenomena occur in 
electronic systems with ``flat-bands'', i.e., 
energy bands with flat dispersions in an entire $k$-space.
Flat-band ferromagnetism is the most well-known example, 
and many research studies have been devoted to this subject.
Mielke first proved the existence of ferromagnetism 
in a Hubbard model on ``line graphs''~\cite{mielke91,mielke92,mielke93}.
Later, Tasaki showed rigorously the existence of 
ferromagnetic ground states in the 
so-called Tasaki model~\cite{tasaki92}.
These tight-binding models 
have a flat-band as the lowest band.
Since the single-electron states
in the lowest band are completely degenerate, the ground states of 
noninteracting models 
with the half-filled lowest band are  highly degenerate 
due to possible spin configurations. The introduction of  
a Hubbard interaction does not change the energy of perfectly
ferromagnetic ground states
since these states do not encounter on-site repulsive 
interactions. If the energies of other states are increased
by such interactions, the so-called flat-band ferromagnetism is realized.
Such a flat-band ferromagnetism can be stable 
against  small perturbations to hopping 
terms~\cite{kusakabe94,tasaki94,tasaki95}.
It is known that a flat-band also appears in a tight-binding 
model on a bipartite lattice with different numbers
of sub-lattice sites. 
The ground state of the half-filled Hubbard model on this type 
of lattice 
was shown to be ferrimagnetic (Lieb theorem)~\cite{lieb89}. 
Motivated by  these studies, several attempts at finding 
a flat-band ferromagnetism in real materials have been carried out. 
Following the guiding concept of constructing localized wave functions, 
tight-binding models with flat-bands
were proposed in several theoretical 
studies~\cite{shima93,fujita96,nishino03,nishino05}. 
For example, in the graphite with zigzag edges, the 
edge states formed a flat-band and the magnetism caused by 
the edge band was predicted~\cite{fujita96}. 

Imada and Kohno argued that 
a flat dispersion may enhance the pairing instabilities
in the systems near a Mott insulator with 
a singlet ground state. According to them, 
the enhanced degeneracy and suppression
of single-particle processes due to the flat dispersion 
may enhance pairing instability in doped systems, 
which may be the cause of the high-$T_c$  in cuprates~\cite{imada99}.

In this manner, the search for tight-binding models 
with flat-bands is strongly desired from 
both theoretical and experimental points of view.
Thus far, several attempts at constructing 
tight-binding models with a flat-band have been carried out.
In most cases, we have to determine 
the parameters in the models as realize a localized wave function
by solving the Schr\"{o}dinger equation directly~\cite{nishino03,nishino05}. 
Although a few systematic methods for constructing flat-band 
tight-binding models,
{\it e.g.}, ``line graph'' and ``cell construction'', 
have been proposed~\cite{mielke91,mielke92,tasaki92}, 
the application of these methods can be possible in 
certain artificial models. Thus, the number of constructed models is limited.
In this paper, we introduce a new simple systematic method
for constructing tight-binding models on a class of lattice 
structures that we call ``partial line graphs''  
in which flat-bands emerge. We can apply this method 
to any two-dimensional (2D) or three-dimensional (3D)
lattices. In addition, the generalization of 
the ``partial line graphs'' is also possible, which
makes the flat-band stable in a wide parameter range. 
Such a flexibility is important as a guideline for constructing 
a material with a flat-band.  
First, we show the method for constructing a partial line graph.
Several tight-binding models produced by generalizing 
the partial line graph also realize 
flat-bands. We describe the rules for the generalization
and show examples of generalized partial line graphs  
of square and cubic lattices.
Finally, we prove the realization of flat-bands 
in these systems.


Given a 2D or 3D lattice,
we can construct its partial line graphs in the following manner.
We divide the lattice into $N$ sublattices ($N>1$), 
where  each site on a sublattice is equivalent and connected 
only to the sites on different sublattices.
Next, we  select a site of one (A) of the sublattices, 
which is connected to $z$ sites on different sublattices, 
and replace the selected site with  $z$ sites 
(A$_1$, A$_2$, $\cdots$, A$_z$) created
on bonds connecting  
the selected site to other sites. We connect 
A$_1$, A$_2$, $\cdots$ and  A$_z$ to each other 
through bonds. Repeating the same process on all the sites on the 
A sublattice,  we construct a partial line graph of the given lattice,
where the A sublattice is replaced by a lattice of clusters composed of
$z$ sites.
Namely, we apply  the concept of the line graph 
only to one of the sublattices.
The constructed lattice has $N+z-1$ sites within a unit cell.
A  tight-binding single-particle model 
with  the same hopping amplitude $t$ on all bonds realizes flat-bands
at the energy $-t$, as will be shown later.

\begin{figure}
\begin{center}
  \includegraphics[]{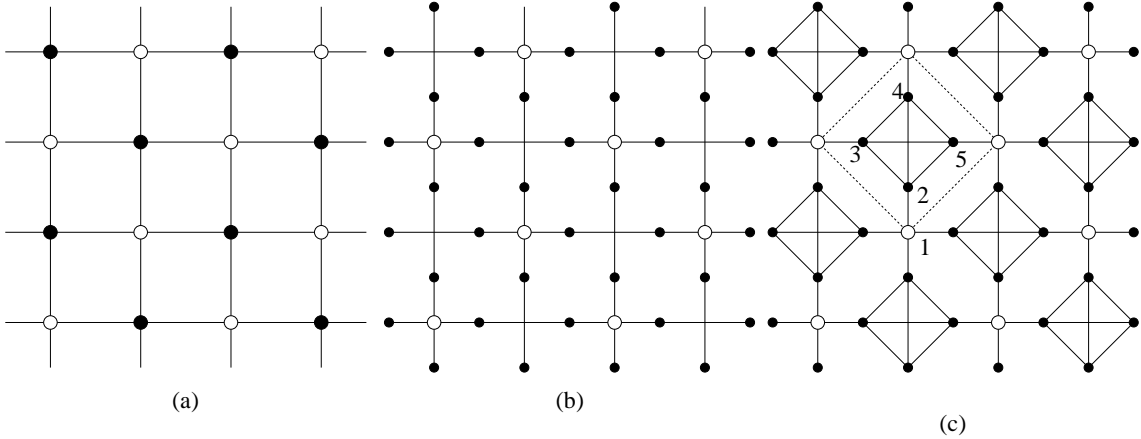}
\end{center}
\caption{(a) 2D square lattice ($z = 4$) with two sublattices, where 
  A and B sublattices are shown by black and white circles, respectively. 
  (b) All sites on the A sublattice are removed,
  and new sites on the bonds connected to the removed sites
  are created.
  (c) The new sites on the clusters are connected. 
  The partial line graph of the 2D square lattice is obtained.
  The unit cell is shown by a dashed line.}
\label{fig:plg-square}
\end{figure}

\begin{figure}
  \begin{center}
    \includegraphics[]{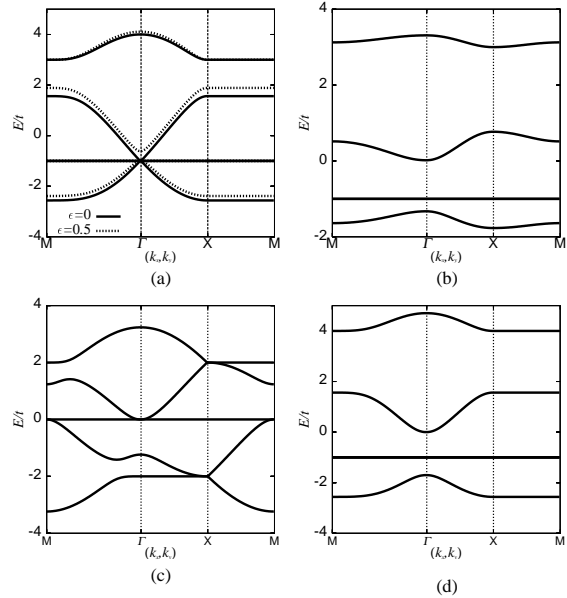}
  \end{center}
  \caption{(a) Dispersion relations for the partial line graph of 
    the square lattice (Fig.~\ref{fig:plg-square}(c)).
    The on-site energy $\epsilon$ on the B site 
    are considered to be
    $0$ and $0.5 t$, while the on-site energies
    within four-site clusters are $0$.
    (b) Dispersion relations for the generalized partial line graph
    shown in Fig.~\ref{fig:generalized-plg}(a) with 
    $t_1/t = 0.2$ and $t_2/t = 0.8$.
    (c) Dispersion relations for the generalized partial line graph
    shown in Fig.~\ref{fig:generalized-plg}(b).
    (d) Dispersion relations for the generalized partial line graph
    shown in Fig.~\ref{fig:generalized-plg}(c).}
  \label{fig:disp_square}
\end{figure}

As an example, let us consider a partial line graph 
of the square lattice. 
First, we divide the lattice into two sublattices 
(A and B), as shown in Fig.~\ref{fig:plg-square}(a). 
We then create four sites (A$_1$, A$_2$, A$_3$ and A$_4$) 
on the bonds connected to the sites on
the A sublattice (see Fig.~\ref{fig:plg-square}(b)).
Connecting the new sites, we obtain 
a partial line graph shown in Fig.~\ref{fig:plg-square}(c),
which contains five unidentical sites in the unit cell. 
Assuming the hopping amplitude $t$ on all bonds,
we calculate the energy dispersion. 
The result is shown in Fig.~\ref{fig:disp_square}(a), 
where doubly degenerate flat-bands exist at $E = -t$.
Furthermore, all the dispersion relations along the highly 
symmetric lines in the ${\bf k}$-space, ${\bf k} = (0, k_y)$ or $(k_x, 0)$, 
are flat, as shown in Fig.~\ref{fig:disp_square}(a).

As mentioned above, we have assumed that all the
hopping amplitudes and
the on-site potential energies on all sites are the same.
However, a flat-band emerges in more generalized models 
containing different hopping amplitudes and/or
on-site energies, 
if we choose these parameters properly.
The generalized partial line graphs are constructed by
the following procedures. In the following discussions, we call 
all the sites other than those on the A sublattice 
simply as B sites, although 
they may belong to different sublattices. 

(i) The on-site energies on the B sites do not affect 
flat-bands at $E = -t$. 
For example,  the value of
the on-site energies of the B sites $\epsilon$
on the partial line graph of the square lattice may be arbitrary.
We show the energy dispersions for $\epsilon = 0.5 t$  
in Fig.~\ref{fig:disp_square}(a) together with those for 
$\epsilon = 0$. 

\begin{figure}
  \begin{center}
    \includegraphics[]{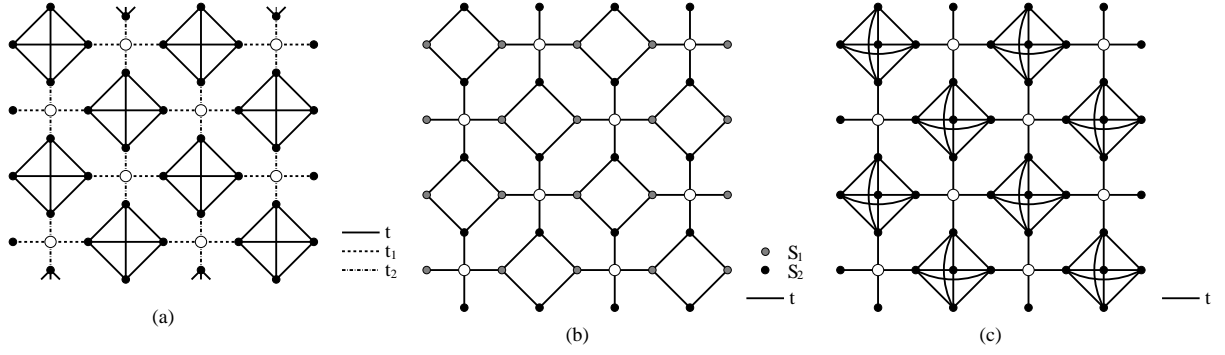}
  \end{center}
  \caption{Generalized partial line graphs of the 2D
    square lattice. 
    (a) Arbitrary hopping amplitudes from
    the B$_1$ original sublattices. 
    (b) Generalization for the hopping amplitude
    within the A cluster. 
    (c) Addition of the original site on the A cluster.}
  \label{fig:generalized-plg}
\end{figure}

(ii) We can take arbitrary hopping amplitudes 
on the bonds that connect the A and B sites or two B sites,
while fixing the amplitudes within A clusters to be $t$.
Flat-bands at $E = -t$ are also realized in this case.
One of these models  
is shown in Fig.~\ref{fig:generalized-plg}(a) and an example
of its band dispersions 
is shown in Fig.~\ref{fig:disp_square}(b).

(iii) The hopping amplitudes within  A clusters can also
be tuned if $z\ge N+2$. We classify $z$ sites
in an A cluster into $M$ subsets 
(S$_1$, S$_2$, $\cdots$, S$_M$), where $M$ should satisfy $z-N \ge M > 1$. 
When the hopping amplitudes between sites in the same 
subset and  between those in 
the subsets $S_i$ and $S_j$ are considered to be $t_0$ and 
$t_{ij}$, respectively, 
the model has a flat-band at $E = -t_0$. 
For example, we divide an A cluster of the partial line graph of the 2D
square lattice into two subsets, $S_1$ and 
$S_2$, as shown in Fig.~\ref{fig:generalized-plg}(b)
and assume $t_0 = 0$ and $t_{12} = t$. 
The band dispersions of this model with a flat-band at $E = 0$
are shown in Fig.~\ref{fig:disp_square}(c).

(iv) We can even maintain the original sites on the A sublattice, 
which was removed in the original partial line graphs. 
This type of lattice contains A clusters composed of $z+1$ sites.
An example of this generalization of the 2D square lattice
and its band dispersions are shown in
Figs.~\ref{fig:generalized-plg}(c) and 
\ref{fig:disp_square}(d), respectively.
Note that applying procedure (iii) for the model 
in Fig.~\ref{fig:generalized-plg}(c) reproduces
the Lieb model~\cite{lieb89}.

\begin{figure}
\begin{center}
  \includegraphics[]{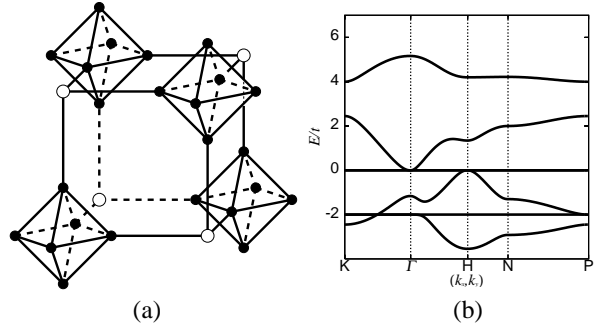}
\end{center}
\caption{(a) Generalized partial line graph of a 3D
  cubic lattice, where all hopping amplitudes on the bonds 
  are assumed to be $t$. (b) Dispersion relation.}
\label{fig:cube}
\end{figure}

Thus far, we have shown only the examples constructed 
from the  2D square lattice. However, many types of tight-binding model 
with flat-bands can be constructed systematically by applying 
the above-mentioned procedures  and their combinations to any type of 
lattice structure.
As an example of 3D models, 
we introduce a generalized partial line graph 
of a 3D cubic lattice (Fig.~\ref{fig:cube}(a)) and show 
its dispersion relations in Fig.~\ref{fig:cube}(b).
In this case, flat-bands appear at $E = 0$ and $-2t$.

Let us now prove
the existence of flat-bands. 
We prove first that flat-bands emerge at $E = -t$ 
on an original partial line graph.
Electronic states are described using the Bloch wave 
function $\phi_{\bf k}$ with the wave vector ${\bf k}$:
\begin{eqnarray}
   \phi_{\bf k} & =\sum_R & \left[ b_{1\,{\bf k}} \phi_{{\rm B}\,1}
   ({\bf R})
    + \cdots + b_{N-1\,{\bf k}} \phi_{{\rm B}\,N-1}({\bf R})
  \right. \nonumber \\
  & & \left.+ a_{1\,{\bf k}} \phi_{{\rm A}\,1}({\bf R})
    + \cdots + a_{z\,{\bf k}} \phi_{{\rm A}\,z}({\bf R})
  \right] e^{i {\bf k} \cdot {\bf R}}, 
  \label{eq:wave}
\end{eqnarray}
where the wave functions $\phi_{{\rm A}\,1}, \cdots, 
\phi_{{\rm A}\,z}$ represent the atomic states within 
 A clusters, and those on
B sites are given by
$\phi_{{\rm B}\,1}, \cdots, \phi_{{\rm B}\,N-1}$.
The wave function  $\phi_{\bf k}$ for the energy $E_{\bf k}$ satisfies the 
eigenvalue equation
${\cal H} \phi_{\bf k} = E_{\bf k} \phi_{\bf k}$.
The Hamiltonian ${\cal H}$ for the wavevector ${\bf k}$
is represented by an $(N+z-1) \times (N+z-1)$ matrix,
which can be written as
\begin{equation}
  {\cal H} =
  \left( \begin{array}{cc}
      \cal{B}({\bf k}) & \cal{G}({\bf k}) \\
      \cal{G}^{\dagger}({\bf k}) & \cal{T}
    \end{array} \right).
  \label{eq:block}
\end{equation}
Here, ${\cal B}({\bf k})$ and ${\cal G}({\bf k})$ are 
the $(N-1) \times (N-1)$ and $(N-1) \times z$ block
matrices, respectively. 
The $z \times z$ block matrix $\cal{T}$ is
independent of the wave vector ${\bf k}$, and all matrix elements,
except the diagonal elements, are $t$:
\begin{equation}
  {\cal T} =
  \left( \begin{array}{cccc}
      0 & t & \cdots & t \\
      t & 0 & \cdots & t \\
      \vdots & \vdots & \ddots & \vdots \\
      t & t & \cdots & 0
    \end{array} \right).
  \label{eq:blocf_Tk}
\end{equation}
Assume that the eigenfunction $\phi_{f\,{\bf k}}$  for the 
eigenvalue  $-t$ has 
no amplitude on B sites, 
{\it i.e.}, $b_{i\,{\bf k}} = 0$ for $1\le i \le N-1$ 
in eq.(\ref{eq:wave}).
Then, all $z$ equations 
$({\cal T} + t I) \ ^{\rm t}(a_{1\,{\bf k}}, a_{2\,{\bf k}}, 
\cdots, a_{z\,{\bf k}}) = 0$
are identical, and the 
eigenvalue equation is reduced to $N$ independent
linear equations for the unknown $z$ coefficients 
$a_{1\,{\bf k}},  a_{2\,{\bf k}}, \cdots, a_{z\,{\bf k}}$:
%
\begin{eqnarray}
\label{eig1}
  {\cal G}({\bf k})\ ^{\rm t}(a_{1\,{\bf k}}, a_{2\,{\bf k}}, 
  \cdots, a_{z\,{\bf k}}) & = & 0, \\
 \label{eig2} 
  a_{1\,{\bf k}} + a_{2\,{\bf k}} + \cdots + a_{z\,{\bf k}} & = & 0.
\end{eqnarray}
Hence, we obtain $z-N$ independent solutions for $z$ unknowns
if the condition $z > N$ is satisfied. That is, we have $z-N$-fold
degenerate energy bands at $E_{\bf k} =-t$, whose dispersions
are flat in the entire ${\bf k}$ space.
For any wave vectors $k$, by making nodes on B sites, 
a wave function takes the eigenvalue $-t$.

It is obvious that the flat-bands at $E = -t$ do not depend on the
on-site energy on the B sites and
the hopping amplitudes from/to the  B sites,  
since they modify only eq.~(\ref{eig1}) and do not change the number of 
independent equations.
Thus, once all the on-site energies in the A clusters 
are identical,
we can choose an on-site energy on the B sites and
an arbitrary amplitude from/to the B sites to realize flat-bands, 
as was already illustrated for the 2D square lattice
 in Figs.~\ref{fig:disp_square}(a) and \ref{fig:disp_square}(b).

In generalization (iii), we tune the hopping amplitudes 
within an  A cluster by dividing the cluster 
into $M$ subsets \{$S_1, S_2, \cdots, S_M$\} 
and taking the hopping amplitudes in the same 
subsets to be $t_0$ and those between 
the subset $S_i$ and $S_j$ to be $t_{ij}$.
We prove that flat-bands emerge at $E = -t_0$.
In this case the block matrix ${\cal T}$ in eq.~(\ref{eq:block})
is written as 
\begin{equation}
  {\cal T} =
  \left( \begin{array}{cccc}
      {\cal T}_0 & {\cal T}_{1\,2} & \cdots & {\cal T}_{1\,M} \\
      {\cal T}_{2\,1} & {\cal T}_0 & \cdots & {\cal T}_{2\,M} \\
      \vdots & \vdots & \ddots & \vdots \\
      {\cal T}_{M\,1}  & {\cal T}_{M\,2} & \cdots & {\cal T}_0
    \end{array} \right),
  \label{eq:T-gen}
\end{equation}
where
\begin{equation}
  {\cal T}_0 =
  \left( \begin{array}{cccc}
      0 & t_0 & \cdots & t_0 \\
      t_0 & 0 & \cdots & t_0 \\
      \vdots & \vdots & \ddots & \vdots \\
      t_0  & t_0 & \cdots & 0
    \end{array} \right),
  \label{eq:T_0}
\end{equation}
and
\begin{equation}
  {\cal T}_{ij} =
  \left( \begin{array}{ccc}
      t_{ij} & \cdots & t_{ij} \\
      \vdots & \ddots & \vdots \\
      t_{ij} & \cdots & t_{ij} 
    \end{array} \right).
  \label{eq:T_ij}
\end{equation}
If we assume the eigenfunction $\phi_{f k}^{\prime}$ with 
no amplitudes on B sites for the eigenvalue
$-t_0$,  the equation
$({\cal T} + t_0 I) (a_{1\,k}, a_{2\,k}, \cdots, a_{z\,k})^t = 0$ 
is transformed to $M$ independent equations:
\begin{eqnarray}
  t_0 A_1 + t_{1\,2} A_2 + \cdots + t_{1\,M} A_M & = & 0 \\
  t_{2\,1} A_1 + t_0 A_2 + \cdots + t_{2\,M} A_M & = & 0 \\
  \vdots \hspace*{2cm}& & \nonumber \\
  t_{M\,1} A_1 + t_{M\,2} A_2 + \cdots + t_0 A_M & = & 0, 
\end{eqnarray}
where
\begin{equation}
  A_m = \sum_{i \in S_m} a_{i\,{\bf k}}.
\end{equation}
By adding eq.~(\ref{eig1}),  we obtain  $N+M-1$ independent equations 
for $z$ unknowns, which leads to 
$z-N-M+1$-fold degenerate flat-bands at $E = -t_0$.
The examples with $t_0 =0$, which result in flat-bands at  $E = 0$, 
are shown in 
Figs.~\ref{fig:generalized-plg}(b) and \ref{fig:cube} (a). 
%
It is interesting to note that in the 3D example (Fig.~\ref{fig:cube}(a)),  
another flat-band appears at $E = -2t$. 
This is a unique feature of the model where 
the hopping amplitudes $t_{ij}$ 
between the subsets $S_i$ and $S_j$ are independent of $i$ and $j$,
and $M=3$.  

Finally, let us prove generalization (iv).
Here, we maintain the original sites on the A sublattice,
which has been eliminated on the partial line graphs.
Even in this lattice, we may assume the wave function $\phi_{f k}$,  
which has no amplitudes on B sites for the eigenvalue $-t$.
Then we have $N$ independent equations for $z+1$ unknowns, and 
can determine the $z-N+1$ independent solutions.
Thus, there are flat-bands at $E = -t$ 
with  $z-N+1$-fold degeneracy. 

In this paper, we have introduced a systematic method 
for constructing
tight-binding models on generalized partial line graphs
with  flat energy bands. 
We have shown above only examples for the  lattices divided originally 
into two sublattices. However, we can apply the method to the lattices 
with any number of sublattices. One realizes that the combinations of
the described generalizations are numerous and useful for
generating flat-band models. 
Since the flatness of the dispersion may lead to
instabilities in the presence of interactions, 
it is important to study  the many-body effects of 
the electrons on the generalized partial line graphs. 
For example, the ground state with a half-filled flat-band
can be a partially polarized ferromagnetic state.
It might be obvious that a ferrimagnetic state is a ground state
in some of the generalized partial line graphs 
of Fig.~\ref{fig:generalized-plg}(c), 
since one of them is the Lieb lattice, where it is proven rigorously
that the ground state is a ferrimagnetic state.
On the other hand, it is well known that 
the fully polarized state on the flat-band $|\Psi\rangle$ is 
the unique ground state, when the flat-band 
is the lowest (highest) energy band and 
the single-particle density matrix 
$\rho_{xy} \propto \langle \Psi | c^{\dagger}_{x \uparrow} 
c_{y \uparrow} |\Psi\rangle$ is irreducible~\cite{mielke93}. 
Although the energy eigenvalue of the flat-band in the partial line graphs 
is not the lowest (highest), such a partially polarized ferromagnetic 
state can be a ground state when the band gaps 
on both sides of the flat-band exist 
for $U/t \ll 1$~\cite{mielke93}. In fact, in  
a chain of square models, where a flat-band can be isolated, 
a flat-band ferromagnetism is realized for small $U/t$
with appropriate parameters~\cite{arita98}.
Since we can isolate a flat-band by tuning the parameters
as shown in Fig2.~\ref{fig:disp_square}(b) and \ref{fig:disp_square}(d), 
a partially polarized ferromagnetic state can be a ground state 
in certain generalized partial line graphs.

We expect that the method is also useful in providing a guideline for
synthesizing materials with flat-bands,
since there is a large room for modifying the models. 
We can vary on-site energies,  hopping amplitudes, and 
other parameters,
while maintaining the flat-bands.
Such flexibility is important in the view point
of material designing. The fact that the on-site energies on B sites 
do not affect the flatness of the dispersion 
is particularly important. This fact indicates that 
the atomic species on B sites 
can be different from those in A clusters in a realistic system.
Thus, we may expect that a  generalized partial line graph
may be realized in certain compounds. 
As a candidate, we expect that the quasi-one-dimensional 
lattice shown in Fig.~\ref{fig:generalized-plg}(a), 
the 2D one in Fig.~\ref{fig:generalized-plg}(b),
and the 3D one in Fig.~\ref{fig:cube}(a)
will be realized in real materials.
It is also useful that
the position of the flat-band can be tuned 
by changing the hopping amplitudes,
which indicates that 
the position of the flat-band 
can be moved toward the Fermi level 
without changing the carrier density 
by applying (chemical) pressure to the material. 
A clusters can be considered as $z$-fold degenerate orbitals
as treated in refs. \citen{nishino03} and \citen{nishino05}, 
which is also useful in
the view point of material designing. 

This work is supported by a Grant-in Aid 
from the Ministry of Education, Culture, 
Sports, Science and Technology of Japan.

\end{document}